                                                                  \documentclass[12pt,preprint]{aastex}

\shorttitle{Unwinding motion of a twisted active-region filament}
\shortauthors{Yan et al.}

\begin{document}

\title{Unwinding motion of a twisted active-region filament}
\author{X. L. Yan\altaffilmark{1,2}, Z. K. Xue\altaffilmark{1}, J. H. Liu\altaffilmark{3}, D. F. Kong\altaffilmark{1}, C. L. Xu\altaffilmark{4}}

\altaffiltext{1}{Yunnan Observatories, Chinese Academy of
              Sciences, Kunming 650011, China.}
\altaffiltext{2}{Key Laboratory of Solar Activity, National Astronomical Observatories, Chinese Academy of Sciences, Beijing 100012, China.}

\altaffiltext{3}{Department of Physics, Shijiazhuang University, Shijiazhuang 050035, China.}

\altaffiltext{4}{Yunnan Normal University, Kunming 650092, China}

\begin{abstract}
To better understand the structures of active-region filaments and the eruption process, we study an active-region filament eruption in active region NOAA 11082 in detail on June 22, 2010. Before the filament eruption, the opposite unidirectional material flows appeared in succession along the spine of the filament. The rising of the filament triggered two B-class flares at the upper part of the filament. As the bright material was injected into the filament from the sites of the flares, the filament exhibited a rapid uplift accompanying the counterclockwise rotation of the filament body. From the expansion of the filament, we can see that the filament is consisted of twisted magnetic field lines. The total twist of the filament is at least 5$\pi$ obtained by using time slice method. According to the morphology change during the filament eruption, it is found that the active-region filament was a twisted flux rope and its unwinding motion was like a solar tornado. We also find that there was a continuous magnetic helicity injection before and during the filament eruption. It is confirmed that magnetic helicity can be transferred  from the photosphere to the filament. Using the extrapolated potential fields, the average decay index of the background magnetic fields over the filament is 0.91. Consequently, these findings imply that the mechanism of solar filament eruption could be due to the kink instability and magnetic helicity accumulation.

\end{abstract}


\section{Introduction}
Magnetic fields of solar prominences/filaments are believed to play an important role in the eruptive flares and coronal mass ejections (CMEs) (Forbes, Priest, \& Isenberg 1994; Antiochos, DeVore, \& Klimchuk 1999; Amari et al. 2000). Moreover, magnetic fields must be capable of supporting against the gravity of the cool dense materials overlying the photosphere neutral line (Tandberg-Hanssen 1995). The investigation on the magnetic fields of solar filaments is still one of the outstanding problems in solar physics.

The structures of solar filaments fall into two broad classes: the sheared three-dimensional arcades (Antiochos et al. 1994; Devore et al. 2000; Aulanier et al. 2002) and the twisted flux ropes (Rust et al. 1996; Aulanier et al. 1998; Amari et al. 1999; van Ballegooijen et al. 2000; Lionello et al. 2002; Yan et al. 2013). Up to now, the nature of solar filament structures is still an open question.

Filament eruptions are due to the non-potential magnetic energy, which has been accumulated in the filament system (Wang et al. 1996). When it reaches a critical point, filaments can not maintain stability and begin to erupt. There are different triggering mechanisms of filament eruptions for energy release, which is described as ideal or resistive process. The idea MHD process can be evidenced by helical kink or torus instabilities of flux ropes or solar filaments (Sakurai et al. 1976; Demoulin \& Priest 1989; Amari et al. 2003; Fan \& Gibson 2003; T$\ddot{o}$r$\ddot{o}$k et al. 2003; Ji et al. 2003; Kliem et al. 2004; Schrijver et al. 2008; Zheng et al. 2014). When the twist of solar filaments or flux ropes reaches a critical value, solar filaments or flux ropes will become unstable (Vr\v{s}nak et al. 1988, 1991; T$\ddot{o}$r$\ddot{o}$k et al. 2004, 2005, 2010; Williams et al. 2005; Green et al. 2007; Srivastava  et al. 2010, Kumar et al. 2012; Yan et al. 2014). The background magnetic field gradient overlying solar filaments or flux ropes plays an important role for torus instability (Kliem et al. 2006; Liu 2008; Guo et al. 2010; Cheng et al. 2013; Zuccarello et al. 2014). The resistive process includes the tether-cutting model and the break-out model (Moore et al. 2001; Antiochos 1998; Shen et al. 2012; Longcope \& Forbes 2014; Sterling et al. 2014). The tether-cutting model requires the reconnection below the filament in a bipolar sheared system and the formed flux ropes or filaments are pulled outward. The break-out model requires the reconnection above the sheared filament channel in a multipolar magnetic field system. With the removing of the overlying magnetic fields, solar filaments will lose their balance and then erupt. Moreover, many simulations were carried out to reproduce the process of these eruption models (Gibson et al. 2006; Fan 2007). In addition, magnetic emergence and cancellation also play important roles in the filament eruptions (Chen et al. 2000; Jing et al. 2004; Jiang et al. 2007; Sterling et al. 2007; Yan et al. 2011).

Recently, rotating magnetic field structures have often been reported by using high resolution data. The tornado-like magnetic structures driven by photospheric vortex flows at the magnetic footpoints are widely named as solar tornadoes. According to the difference in spatial and temporal scales and whether the rotating magnetic field structures are connected to prominences or not, solar tornadoes can be divided into two different types: one type is the supposed rotating legs of prominences (Li et al. 2012; Su et al. 2012; Wedemeyer et al. 2013; Panesar et al. 2013); The other is the small-scale solar tornadoes, which are observed on-disk in connection with small-scale vortex flows and swirling plasma motions without any confirmed connection to prominences (Wedemeyer-B\"{o}hm et al. 2012). The observations and three-dimensional numerical simulations found that the small magnetic tornadoes on-disk are caused by the small-scale vortex and swirling plasma motions in the photosphere, which provide the energy from the lower atmosphere to the upper atmosphere (Wedemeyer-B\"{o}hm et al. 2012;Wedemeyer \& Steiner 2014). Moreover, the former type is larger than the latter one in size (Wedemeyer et al. 2013). The interesting issue is whether the magnetic structures in tornado-like prominences are indeed rotating and what cause the rotation of these magnetic structures. Spectroscopic observations of German Vacuum Tower Telescope and Hinode/EIS found the existence of the foot rotation of quiescent prominences (Su et al. 2014; Orozco Su$\acute{a}$rez et al. 2012). However, Panasenco et al. (2014) found that solar tornadoes relative to prominences reported by previous researchers may be due to the counter-streaming and oscillations or the projection on the plane of the sky of plasma motion along magnetic field lines, rather than from a true vortical motion around a vertical or horizontal axis. For the rotating magnetic structures without any connection to prominences, Zhang \& Liu (2011) and Yu et al. (2014) found the ubiquitous rotating network magnetic fields and EUV cyclones on the solar disk and rooted in the rotating network magnetic fields. Up to now, the nature of solar filament-related tornadoes is still controversial.

In order to address the structure of solar filaments and the nature of solar tornadoes relative to the filaments, we present a solar tornado-like filament eruption observed by SDO on June 22, 2010. The detailed observations and methods are presented in section 2. The results are shown in section 3. The conclusion and the discussion are given in section 4.

\section{Observation and method}
The instruments onboard Solar Dynamics Observatory (SDO) can provide simultaneously multiple wavelength observation of solar atmosphere. The Atmospheric Imaging Assembly (AIA; Lemen et al. 2012) onboard SDO provides high-resolution full-disk images of the transition region and the corona with a spatial resolution of 1.$^\prime$$^\prime$5 and a cadence of 12 s. The AIA can observe 10 narrow UV and EUV passbands. The individual AIA channels are sensitive to a wide range in temperature from 6 $\times$ 10$^4$ K to 2 $\times$ 10$^7$ K. The field of view of AIA can be extended to 1.3 solar radius.
From the high spatial and temporal resolution data of SDO, the detailed evolution and eruption process of solar activities can be seen clearly.
One of these wavelengths, 304 \AA\ (He II, T = 0.05 MK), is very suitable to trace the erupting process of solar filaments.
Moreover, Helioseismic and Magnetic Imager (HMI; Schou et al. 2012) onboard SDO can provide full-disk line-of-sight magnetograms at 45 s cadence with a precision of 10 G.

In this paper, we mainly use 304 \AA\ images to study the evolution of a twisted active-region filament and the line-of-sight magnetograms to calculate the helicity of the whole active region and the foot of the filament. All data were calibrated with standard Solarsoft routines and all images observed by SDO are differentially rotated to a reference time (08:00:02 UT).

The differential affine velocity estimator (DAVE; Schuck 2006) is used to derive the photospheric flow field from HMI line-of-sight magnetograms. The transport rate of magnetic helicity from the sub-photosphere to the corona by the photospheric horizontal motions is described by the following equation:

\begin{equation}
\frac{dH}{dt}=-2\oint(A_p \cdot u)B_n{d^2}x,
\end{equation}
where $A_p$ is the vector potential of the potential field, $u$ is the photospheric transverse velocity computed by the DAVE method, $B_n$ is the normal component of the magnetic field. We have computed the transport rate of helicity following the method of Chae et al. (2001).  The DAVE method combines the advection equation and a differential feature tracking technique to detect flow fields. We used a window size of 19 pixels according to the former studies (Liu et al. 2013), which is large enough to include structure information and small enough to have a good spatial resolution. The window size is  determined by examining the slope, Pearson linear correlation coefficient, and Spearman rank order (see Schuck 2008 for discussion of window selection in section 3).

\section{Result}
\subsection{Material motion before the filament eruption}
Figure 1 shows a 304 \AA\ image observed by AIA onboard SDO at 08:00:02 UT on June 22, 2010. An active-region filament is marked by the white box in active region NOAA 11082. Moreover, the white box indicates the field of view of Fig. 2 and Fig. 4. The black box indicates the field of view of Fig. 3 and Fig. 5. The green box indicates the region that is used to calculate the magnetic helicity at the filament foot. The shape of the filament looks like a ponytail. There is an obvious unidirectional material motion in the filament from 08:03:02 UT to 08:34:14 UT. Sequence of 304 \AA\ images in Fig. 2 are used to show the motion of the filament material. The bright material marked by the white arrows in Figs. 2(a)-2(b) can be seen from the lower part to the upper one. In the following, the other two parts of the dark materials marked by the white arrows in Figs. 2(c)-2(d) and Figs. 2(e)-2(f) shifted successively from the lower part to the upper one. Two small brightening patches appeared at the upper foot of the filament at 08:41:50 UT. Next, the bright material marked by the white arrows in Figs. 2(g)-2(h) can be found to shift from the upper part to the lower one. Note that the white arrows in Fig. 2(a) and Fig. 2(b) denote the same feature as well as that in Fig. 2(c) and Fig. 2(d), Fig. 2(e) and 2(f), Fig. 2(g) and Fig. 2(h).
\subsection{The unwinding motion of the filament}
As the bright material was shifting from the upper part to the lower part of the filament, the lower part of the filament became activated and began to exhibit a counterclockwise rotation at 08:51:38 UT. Sequence of 304 \AA\ images in Fig. 3 are used to show the rotation of the filament. Note that the field of view of Fig. 3 is marked by the black box in Fig. 1. In order to show the rotation of the filament body, we choose four examples to show the rotation of the filament body by tracing the motion of the dark material. The white dashed lines outline the structure of the filament with time. From 09:01:14 UT to 09:04:38 UT, the dark material marked by the white arrows in Figs. 3(a)-3(b) can be seen to shift from one side to the other side of the filament. The other three examples are shown in Figs. 3(c)-3(d), 3(e)-3(f), and 3(g)-3(h). The arrows in Figs. 3(a) and 3(b) denote the same feature as well as that in 3(c) and 3(d), 3(e) and 3(f), 3(g) and 3(h). Because the selected dark features are very easy to identify from their evolution, we selected four examples from these features to show the material movement in the filament. The filament eruption resulted in a simultaneous rotation of the filament from 08:53:38 UT to 10:37:26 UT. During the filament eruption, two B-class flares were observed at the upper part of the filament. The B1.0 flare began at 09:00 UT, peaked at  09:04 UT, and ended at 09:06 UT (Figs. 4(a)-4(c)). The B2.1 flare began at 09:33 UT, peaked at  09:38 UT, and ended at 09:41 UT (Figs. 4(d)-4(f)).  After the occurrence of two flares, the bright materials can be seen to be injected into the lower part of the filament. The flares occurred during the filament eruption. Furthermore, the flares provided the hot materials for the filament and caused the rapid uplift of the filament. The magnetic field lines winding around the filament axis expanded and experienced untwisting motion during the filament eruption. Obviously, the rotation of the filament body was the unwinding motion of the twisted magnetic field lines. The unwinding motion resembled a tornado although the underlying physical mechanism is distinct from the events described by Su et al. (2012) and Wedemeyer et al. (2013). The observed rotation reported here is caused by the unwinding of a previously twisted magnetic field structures not due to vortex flows at the photospheric footprints of the magnetic fields.

Sequence of 304 \AA\ running difference images from 09:39:50 UT to 10:11:02 UT are used to show the process of the rapid expansion and rotation during the filament eruption (Fig. 5). The rising of the filament was accompanied with the body rotation. The process of the filament eruption was like a solar tornado, which was caused by the unwinding motion of the filament. After the unwinding motion, the filament disappeared. Figure 6 shows 304 \AA\ image at 09:59:04 UT with the velocity map superimposed. The velocity field was calculated by using LCT method and 304 \AA\ images (November \& Simon 1988; Chae 2001). The LCT method determines the flow fields by estimating the displacement of a rigid subregion between two consecutive images. The similarity between two subregions can be estimated from the correlation function. The local velocity is determined by estimating a local maximum value in the correlation function with respect to the velocity vector (November \& Simon 1988). The blue arrows in Fig. 6 denote the flow fields. The maximum velocity is 14 $\pm$ 2.7 km/s and the time period of the flow fields is one minute from 09:58:04 UT to 09:59:04 UT. The white box in Fig. 6 has the same field of view as the black box in Fig. 1. The spiral morphology of the flow field can be seen from the upper part of the filament and the material flows were drained from the lower part of the filament.

During the filament eruption, the body of the filament exhibited a significant counterclockwise rotation. The rotation of the filament was measured as horizontal movement at two positions by using time slice method. The field of view of Fig. 7(a) is the same as that of Fig. 1. Figure 7(b) and 7(c) give two time slices acquired at the positions marked by s1 and s2 in Fig. 7(a). The rotation of the dark and the white features can be seen clearly from the time slices. By tracing the dark features of the filament, the rotation angle of the features can be estimated. In order to identify the truth of the rotation angle, we traced carefully the evolution of the filament from one image to the next. The rotation angle obtained by using time slice method is indeed caused by the rotation of the filament not due to the oscillations or other dynamic effects. Because the dark features are clear to identify during their evolution, we chose these features as the tracer. The red arrows in Fig. 7(b) indicate the dark features moved from one side of the filament to the other. There are ten rotational dark features marked in Fig. 7(b). Therefore, the rotation angle is at least 5$\pi$. Fig. 7(c) just shows the rotation of the upper part of the filament. The rotational angle is about 1.5$\pi$. The rotation angle was calculated just by tracing the dark features. The white dashed lines in Figs. 7(b) and 7(c) indicate the boundary of the filament width with time. The width of the filament at the position s1 increased during the filament eruption. The morphology of the filament exhibited a funnel shape during the unwinding motion. Moreover, it implies that the filament was a twisted flux rope before the eruption.

\subsection{The decay index}
Except for kink instability, torus instability is also believed as one of the trigger mechanisms for filament eruptions (Schrijver et al. 2008; Kliem et al. 2014; Zuccarello et al. 2014). The occurrence of torus instability depends on the decay index of the background magnetic fields over filaments. The decay index is defined as follows (Kliem \& T$\ddot{o}$r$\ddot{o}$k 2006; Liu 2008; Xu et al. 2012):

\begin{equation}
n=-\frac{d log(B_t)}{d log(h)},
\end{equation}

where $B_t$ is the strength of the background magnetic fields in the transverse direction and $h$ is the radial height above the photosphere. The background magnetic fields are computed from the line-of-sight HMI magnetogram over the solar surface based on a potential field source surface model (Schrijver \& De Rosa 2003). The method assumes that the coronal and inner-heliospheric magnetic fields are approximatly potential fields. The magnetic fields measured at the photosphere (located at r = 1) were taken as the initial condition to extrapolate the coronal magnetic fields (see Appendix C in Schrijver \& De Rosa 2003). From the evolution of the magetograms, the change of this active region was not very rapid. The extrapolation may approximatly denote the background magnetic fields before the filament eruption. Figure 8 shows the magnetogram at 08:00:00 UT overplotted with the extrapolated potential field lines (red lines). The green line indicates the polarity inversion line (PIL) below the active-region filament. We choose 10 points (marked by the blue asterisks) along the PIL from the upper to the lower to calculate the decay index. The upper panel of Fig. 9 shows $log(B_t)$ versus $log(h)$ along the PIL. The height range of 46.5 - 114.4 Mm was used for the extrapolated potential fields. The background magnetic fields were believed to dominate in the range of about 40 - 100 Mm (Liu 2008; Xu et al. 2012). The lower panel of Figure 9 shows the decay indices derived from the extrapolated potential field along the PIL from the upper part to the lower one. The maximum value of the decay index is less than 1. On average, the decay index along the PIL is 0.91. Previous theoretical works found that the critical decay index ranges from 1.5 to 2.0 for torus instability (T$\ddot{o}$r$\ddot{o}$k \& Kliem 2005; Kliem \& T$\ddot{o}$r$\ddot{o}$k 2006). Liu (2008) investigated several events and found that if the decay index is lower than 1.71, the filament cannot erupt successfully. The decay index for this event is lower than the critical value for torus instability.

\subsection{Transport of magnetic helicity}
Magnetic helicity injection is also important for solar eruptions. In order to address this issue, we calculate the helicity rate and the helicity accumulation before and during the filament eruption. We use the line-of-sight HMI magnetograms to calculate the magnetic helicity injection rate and accumulation. Sequence of line-of-sight HMI magnetogram in Fig. 10 are used to show the evolution of the active region during the filament eruption. The green box is the same as the green box in Fig. 1, which is used to calculate the magnetic helicity at the foot-point of the filament. The dashed lines in Fig. 10a outline the boundary of the filament. We calculate the helicity injection rate and the magnetic accumulation of the whole active-region and the green box in Fig. 10, respectively. Figure 11 shows the temporal variation of the magnetic helicity injection rate and magnetic helicity accumulation for the whole active-region (Figs. 11(a) and 11(b)) and the foot-point of the filament (Figs. 11(c) and 11(d)), respectively. Both magnetic helicity injection rate and the magnetic accumulation are negative before and during the filament eruption. As the unwinding motion of the filament was counterclockwise, the helicity of the filament can be induced to be negative. According to the change of the helicity rate and helicity accumulation, the continuous magnetic helicity was injected into the filament from the foot-point of the filament before and during the filament eruption. The calculation of the helicity injection rate and the magnetic accumulation depends on the velocity calculation. The standard error of the velocity calculation is about 15\% (estimating from the calculated velocity). This implies that the non-potentiality energy was transferred from the lower atmosphere to the upper atmosphere.

\section{Conclusion and discussion}
In this paper, we use the high resolution data observed by SDO and address three questions: One is that the structure of the active-region filament is a twisted flux rope. The second is that the tornado-like motion is caused by the unwinding motion of the filament. The third is that the magnetic helicity can be transferred from the photosphere to the corona. In addition, we find that the total twist in the filament is at least 5$\pi$ and there is a continuous magnetic helicity injection from the photosphere. Moreover, the magnetic fields above the filament are not satisfied with the occurrence condition of the torus instability. Consequently, the mechanism of the filament eruption may be kink instability and magnetic helicity accumulation.

Active-region filaments are different from quiescent filaments (Martin 1998; Zhou et al. 2014). Quiescent prominences have spines and barbs, while active-region filaments have not barbs. Up to now, two popular viewpoints for the structure of solar filaments are the arcade model and the twisting flux rope model. The exact structures of solar filaments need 3D magnetic field measurement. Due to the 2D high resolution data, it is very difficult to know the structures of solar filaments. Especially, active-region filaments have small size compared with quiescent filaments. Even though high resolution data are obtained by the ground based telescopes, the exact magnetic structures of the active-region filament have not been observed. A lot of MHD simulations were carried out to recurrent the process of solar eruptions (Gibson et al. 2006; Fan 2010; Murawski et al. 2014; Zaqarashvili et al. 2014). These simulations assume the existence of the flux rope in solar atmosphere. By tracing the material motion, Li et al. (2013), Li \& Zhang (2013), and Yang et al. (2014) identify the existence of flux rope. However, these observations have never found high twisted flux ropes. Fortunately, one can detect the structures of the filaments from the evolution by using high temporal and spatial data. The event studied in this paper provides a good opportunity to detect the fine structure of the active-region filament. During its eruption, the filament expanded, accompanied with the strong unwinding motion. The magnetic field lines can be seen unwinding around the axis of the filament. It implies that the active-region filament is a very twisted flux rope.

The MHD helical kink instability and the torus instability have been suggested as the trigger and initial driving mechanisms for solar filament eruptions or coronal magnetic flux ropes. Hood \& Priest (1979, 1981) used a numerical method to investigate the stability of cylindrically symmetric magnetic fields. The loop is unstable if the twist is greater than 2.49$\pi$. The simulation of T\"{o}r\"{o}k \& Kliem (2003) found that if a critical end-to-end twist of the filament is larger than 2.75$\pi$, the filament will become unstable and erupt. There are some observational evidence to confirm the existence of kink instability (Srivastava et al. 2010; Kumar et al. 2012; Yan et al. 2014).  The on-disc event studied in this paper exhibited apparent unwinding rotation during the filament eruption. The unwinding motion implies that the magnetic field lines  were wrapping the axis of the filament. Through the unwinding motion, the twist in the filament can be derived roughly. The twist in this active-region filament reached at least 5$\pi$, which is greater than 2.75$\pi$. According to the kink instability condition of the theoretical and simulation study, the twist derived from the unwinding motion of the filament leg is larger than the critical value obtained by previous authors (Hood \& Priest 1981; T$\ddot{o}$r$\ddot{o}$k et al. 2003). The filament eruption may be due to kink instability. In fact, the twist of this filament was underestimated, because we just measured the twist from the rotation of the filament. Even so, the filament still satisfied the condition of kink instability. We also calculate the decay index of the background magnetic fields over the filaments. The maximum value of the decay index is less than 1 and lower than the critical value for the occurrence of torus instability obtained by some previous researchers (T$\ddot{o}$r$\ddot{o}$k \& Kliem 2005; Kliem \& T$\ddot{o}$r$\ddot{o}$k 2006; Liu 2008). They obtained the critical decay index ranges from 1.5 to 2.0 for torus instability.  Therefore, torus instability  can be excluded from the trigger mechanisms of this active-region filament.

Tornado-like prominences were observed for decades (Vr\v{s}nak et al. 1980; Liggett \& Zirin 1984; Wang et al. 1996). Following high-cadence EUV imagery of SDO/AIA observation, more new details of tornado-like prominences were found (Li et al. 2012; Su et al. 2012; Wedemeyer et al. 2013; Wedemeyer et al. 2013). Li et al. (2012) found that the rotation of the filament legs can derive the material from the lower atmosphere. The observation and simulation of Wedemeyer-B\"{o}hm et al. (2012) found that the small-scale magnetic tornadoes without any connection to prominences are driven by photospheric vortex flows. However, for the event studied in this paper, it is very clear that the tornado-like filament eruption formed due to the unwinding motion of the filament.

Magnetic helicity plays an important role in solar eruptions (Romano et al. 2009; Ravindra et al. 2011; Vemareddy et al. 2012; Park et al. 2013; Thompson 2013; Dhara et al. 2014). From the calculated helicity before and during the filament eruption, we find that the magnetic helicity increased continuously. It implies that the energy can be transferred from the photosphere to the corona. The symbol of the magnetic helicity is negative. The filament have left-hand twist. The chirality of them is consistent with each other. The process is very like the sunspot rotation, which can twist the magnetic field lines and the twist can be transferred from the photosphere to the corona (Yan et al. 2012; Ruan et al. 2014). Similarly, Wedemeyer-B\"{o}hm et al. (2012) and Yan et al. (2013) found that the energy and twist can be transmitted by vortex motion at the magnetic foot-points on small spatial scales from the inner atmosphere to the outer solar atmosphere via magnetic fields. Through the calculation of magnetic helicity of this active region, it confirms that the magnetic helicity can be transferred from the photosphere to the corona. This filament eruption is also due to the accumulation of the magnetic helicity.

\acknowledgments
We thank the referee for his/her constructive suggestions and comments that helped to improve this paper. SDO is a mission of NASA's Living With a Star Program. The authors are indebted to the SDO team for providing the data. This work is supported by the National Science Foundation of China (NSFC) under grant numbers 11373066, 1178016, HZKT201309, XJPT002, Yunnan Science Foundation of China under number 2013FB086, 2013FZ041, the Talent Project of Western Light of Chinese Academy of Sciences, Key Laboratory of Solar Activity of CAS under number KLSA201303, KLSA201212, and HeBei Natural Science Foundation of China under number A2010001942.

\begin{figure}
\epsscale{.80}
\plotone{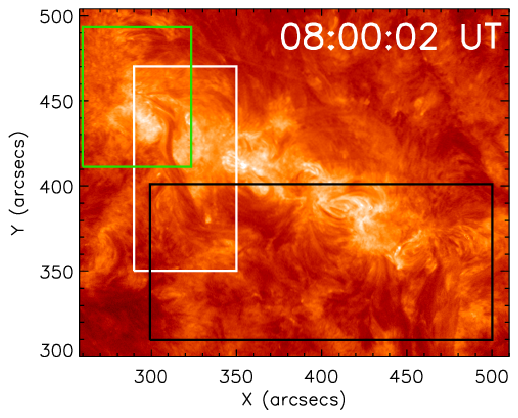}
\caption{ 304 \AA\ images at 08:00:02 UT observed by SDO. The white box indicates the active-region filament and the field of view of Figs. 2 and 4. The black box indicates the field of view of Figs. 3 and 5. The green box indicates the region that is used to calculate the magnetic helicity at the filament foot. \label{fig1}}
\end{figure}

\begin{figure}
\epsscale{.90}
\plotone{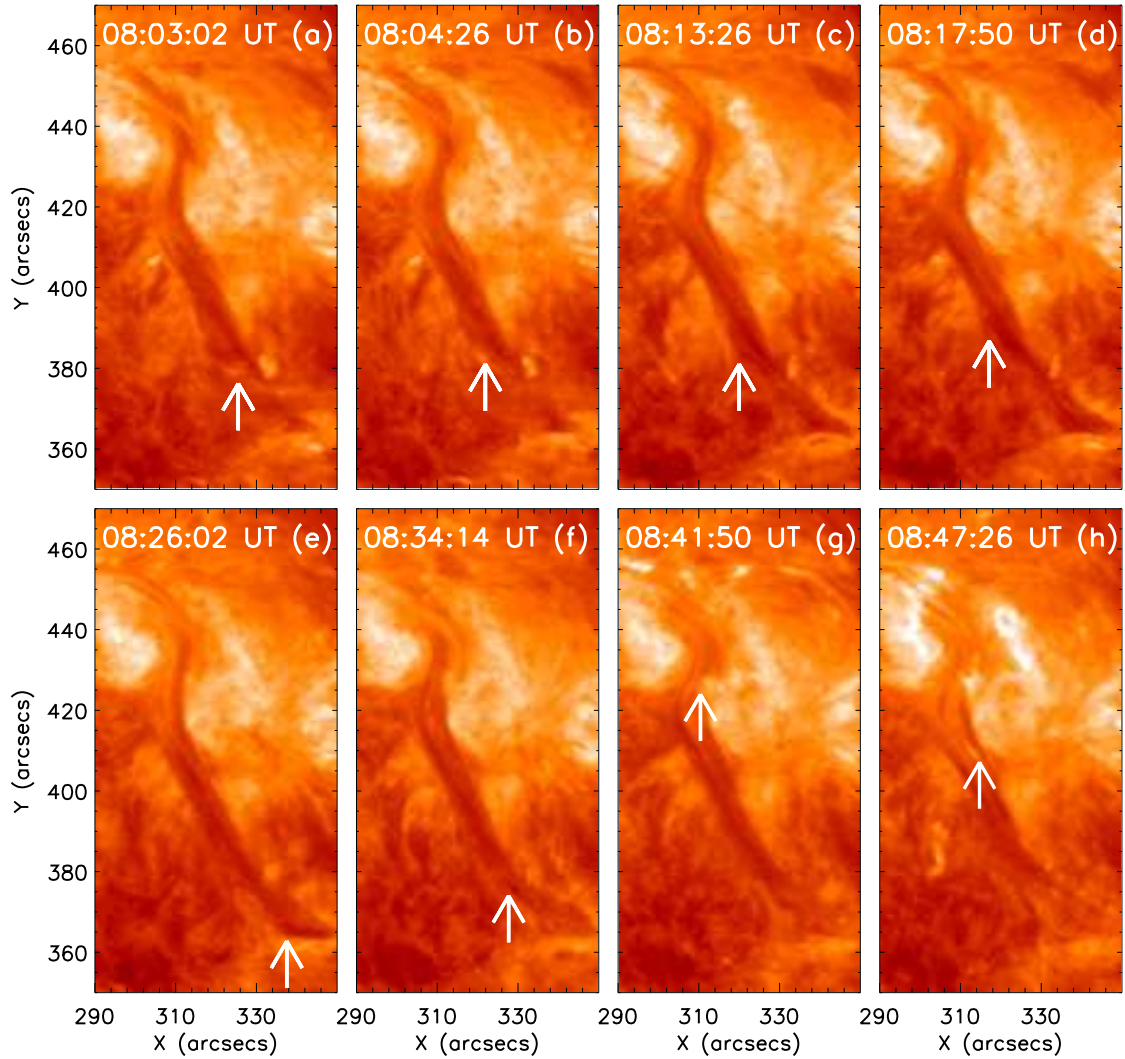}
\caption{Sequence of 304 \AA\ images to show the material motion of the active-region filament from 08:03:02 UT to 08:47:26 UT on 2012 June 22. The white arrows in Figs. 2(a)-2(f) denote the positions of the filament materials that moved successively from the lower part to the upper one. The white arrows in Figs. 2(g)-2(h) denote the positions of the filament materials that moved from the upper part to the lower one. Note that the white arrows in Figs. 2(a) and 2(b) denote the same feature as well as that in Figs. 2(c) and 2(d), Figs. 2(e) and 2(f), Figs. 2(g) and 2(h).\label{fig2}}
\end{figure}

\begin{figure}
\epsscale{.90}
\plotone{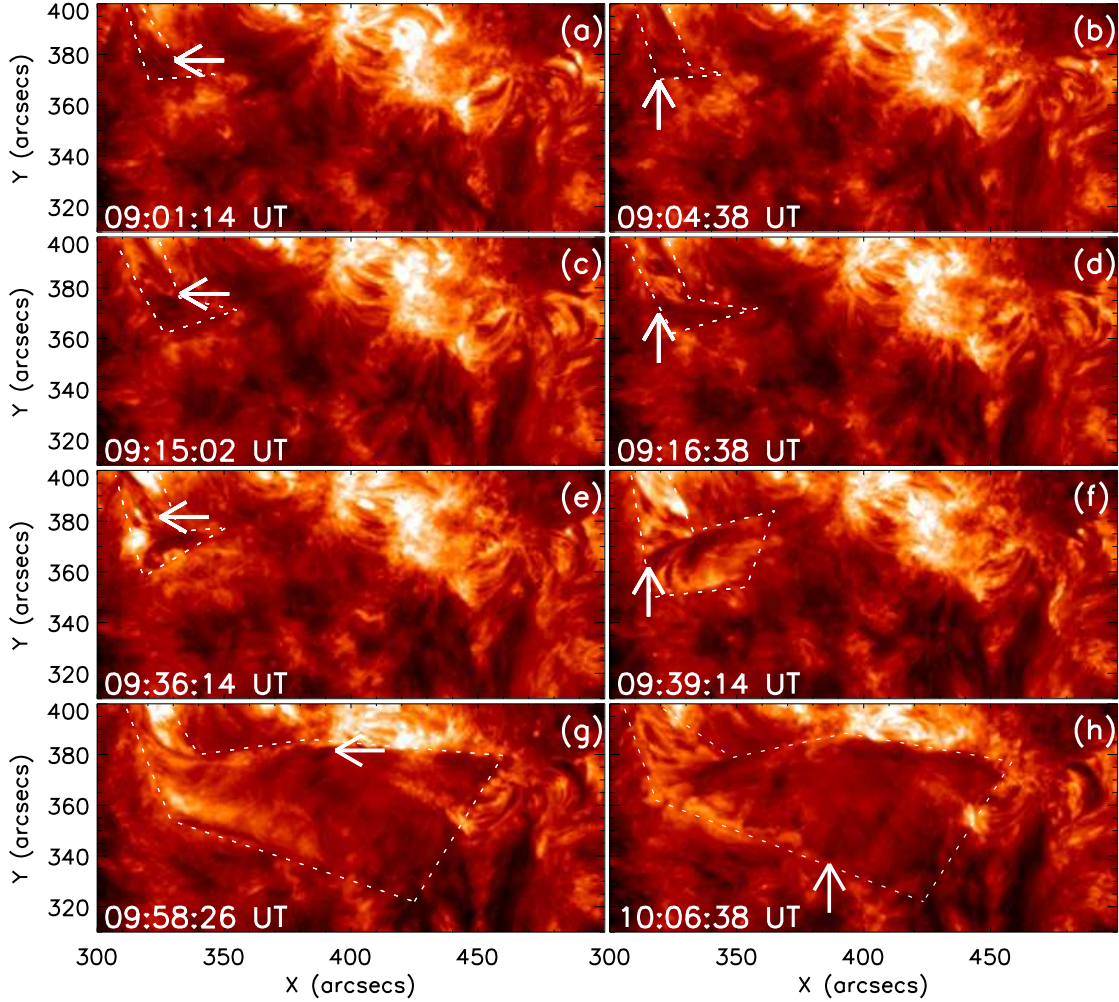}
\caption{Sequence of 304 \AA\ images to show the rotation process of the active-region filament from 09:01:14 UT to 10:06:38 UT on 2012 June 22. The white arrows denote the features that rotated from left to right of the active-region filament. Note that the field of view of the 304 \AA\ images  is marked by the black box in Fig. 1. The white dashed lines outline the structure of the filament with time. Note that the arrows in Figs. 3(a) and 3(b) denote the same feature as well as that in 3(c) and 3(d), 3(e) and 3(f), 3(g) and 3(h). \label{fig2}}
\end{figure}

\begin{figure}
\epsscale{.90}
\plotone{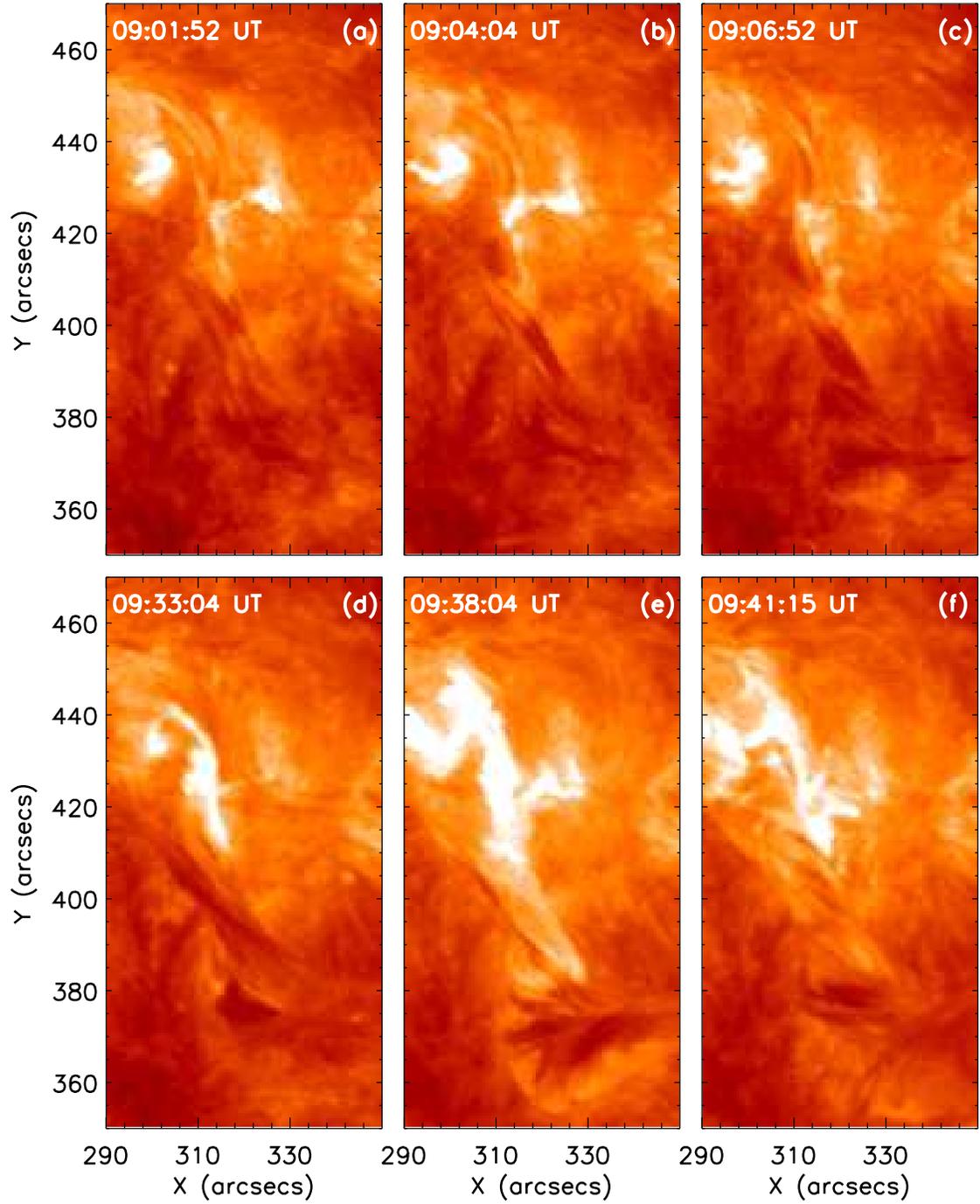}
\caption{Sequence of 304 \AA\ images to show two B-class flares acquired at 304 \AA\ images from 09:01:52 UT to 09:41:15 UT on 2012 June 22. The field of view is the same as that of Fig. 2.\label{fig2}}
\end{figure}

\begin{figure}
\epsscale{.90}
\plotone{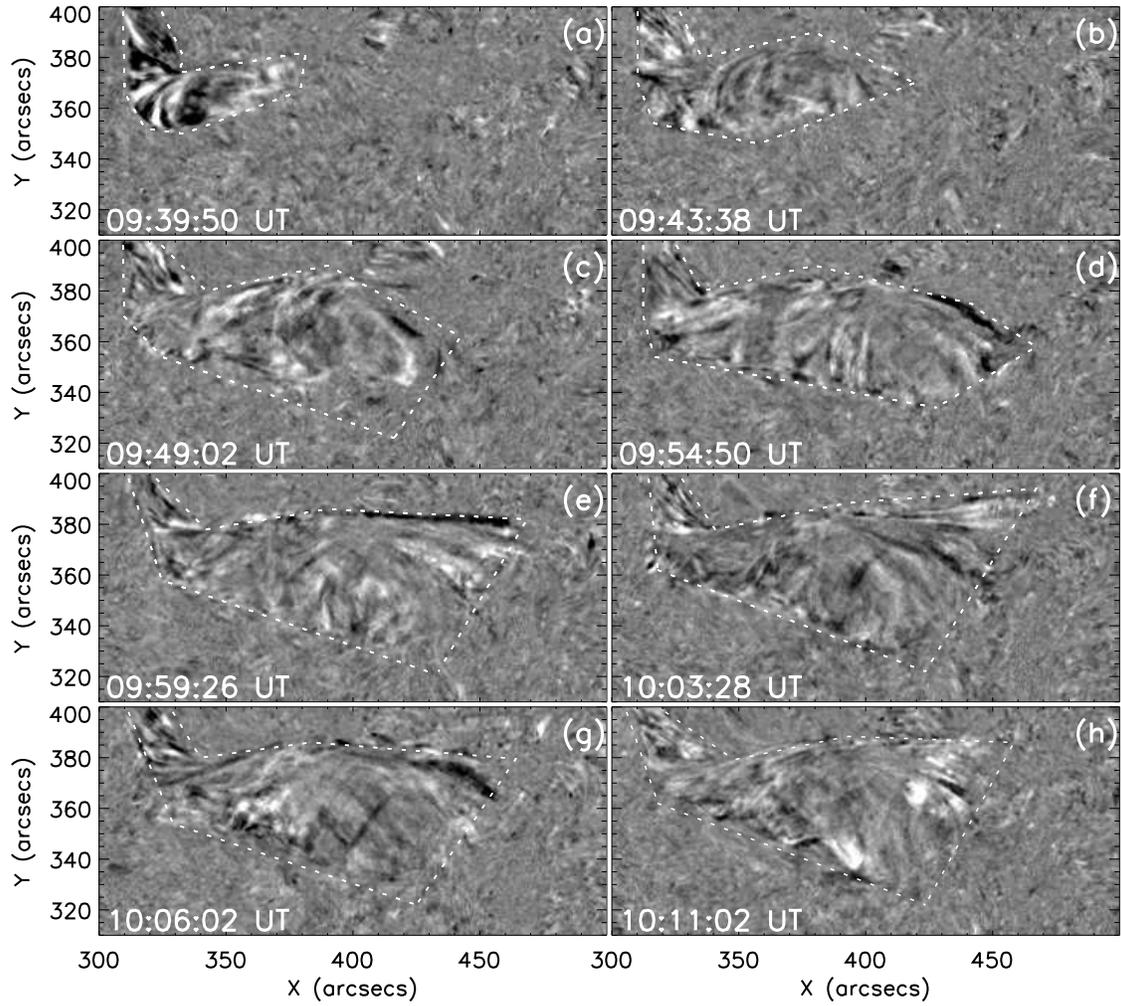}
\caption{Sequence of 304 \AA\ running-difference images to show the eruption process
of the active-region filament from 09:39:50 UT to 10:11:02 UT on 2012 June 22. The field of view is the same as that of Fig. 3. The white dashed lines outline the structure of the filament with time.\label{fig2}}
\end{figure}

\begin{figure}
\epsscale{.90}
\plotone{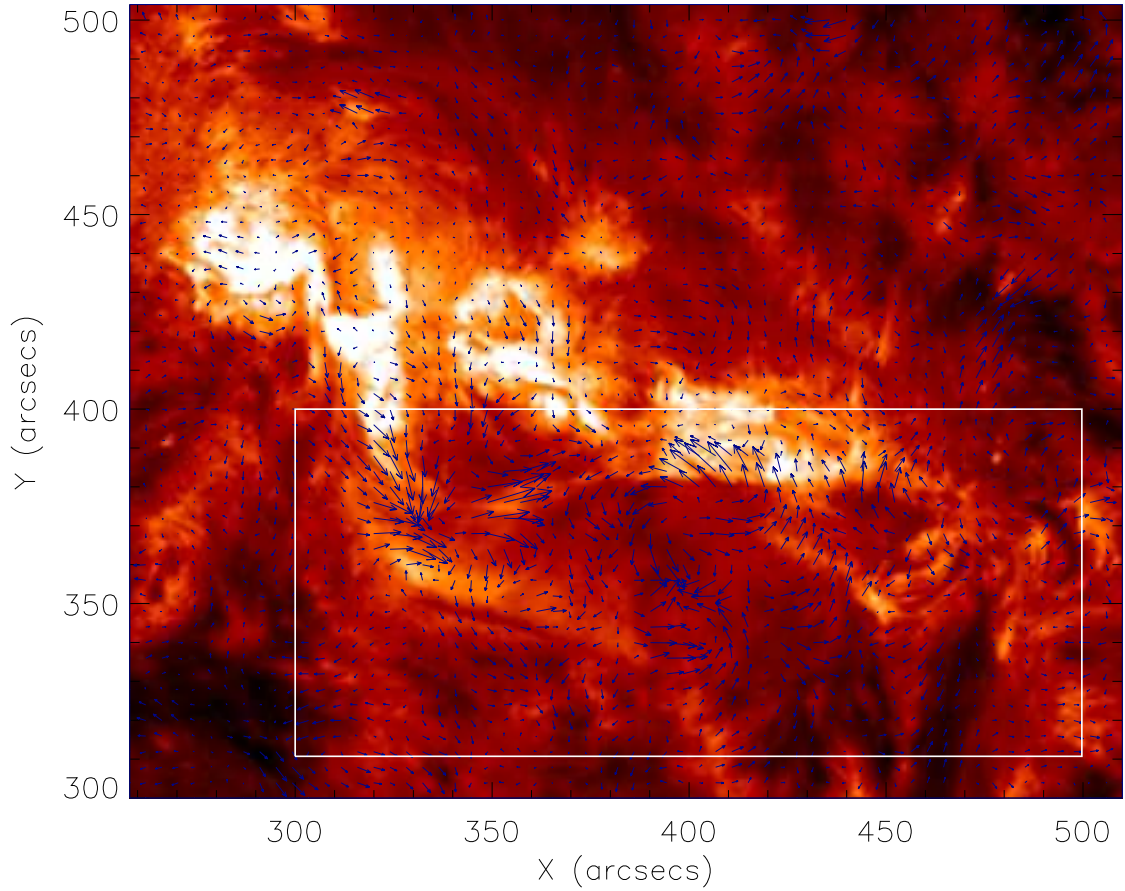}
\caption{304 \AA\ image taken at 09:59:04 UT during the filament eruption, over-plotted with the flow fields calculated by the LCT method. The time period of the flow fields is one minute from 09:58:04 UT to 09:59:04 UT.\label{fig2}}
\end{figure}

\begin{figure}
\centering
\includegraphics[angle=0,scale=2.2]{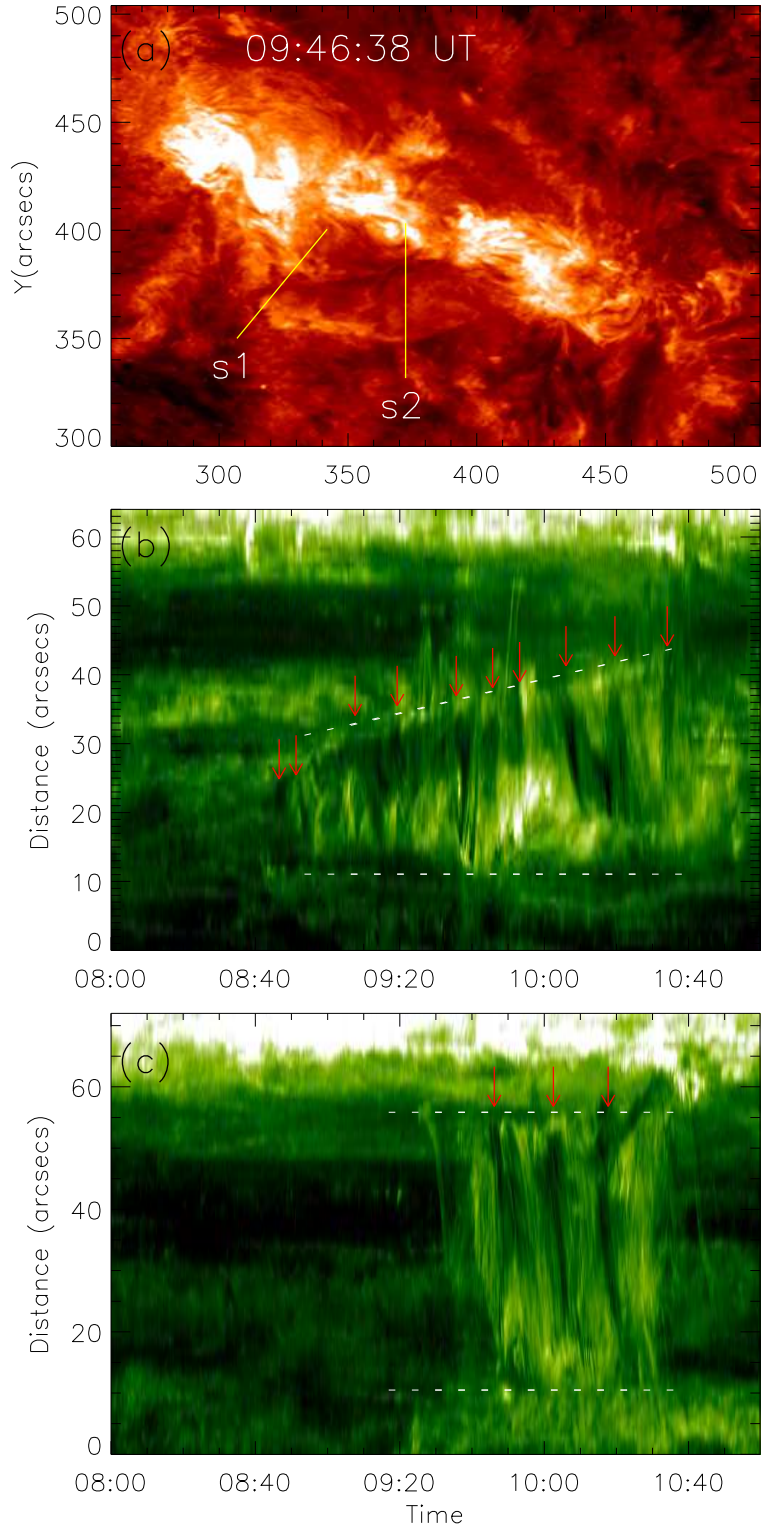}
\caption{Time-slices taken from the 304 \AA\ images at the position marked by the yellow lines in Fig. 7(a), respectively. The yellow lines s1 and s2 in Fig. 7(a) show the positions of the time-slices of Fig. 7(b) and Fig. 7(c), respectively.  The red arrows in Figs. 7(b) and 7(c) indicate the dark features that moved from one side of the filament  to the other side. The field of view of Fig. 7(a) is the same as that of Fig. 1.}
\end{figure}

\begin{figure}
\epsscale{.80}
\plotone{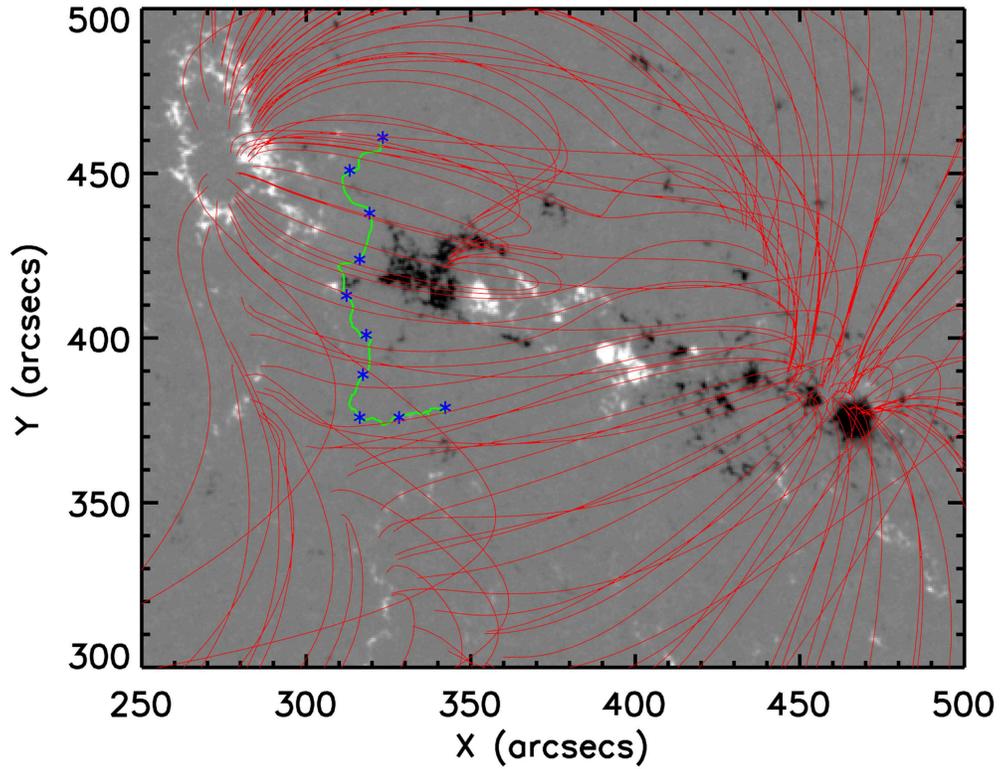}
\caption{ Line-of-sight HMI magnetogram taken at 08:00:00 UT before the filament eruption, overplotted with the extrapolated potential field lines (red). The green line indicates the polarity inversion line (PIL). The blue asterisks indicate the positions used to calculate the decay index along the PIL.\label{fig1}}
\end{figure}

\begin{figure}
\centering
\includegraphics[angle=0,scale=1.5]{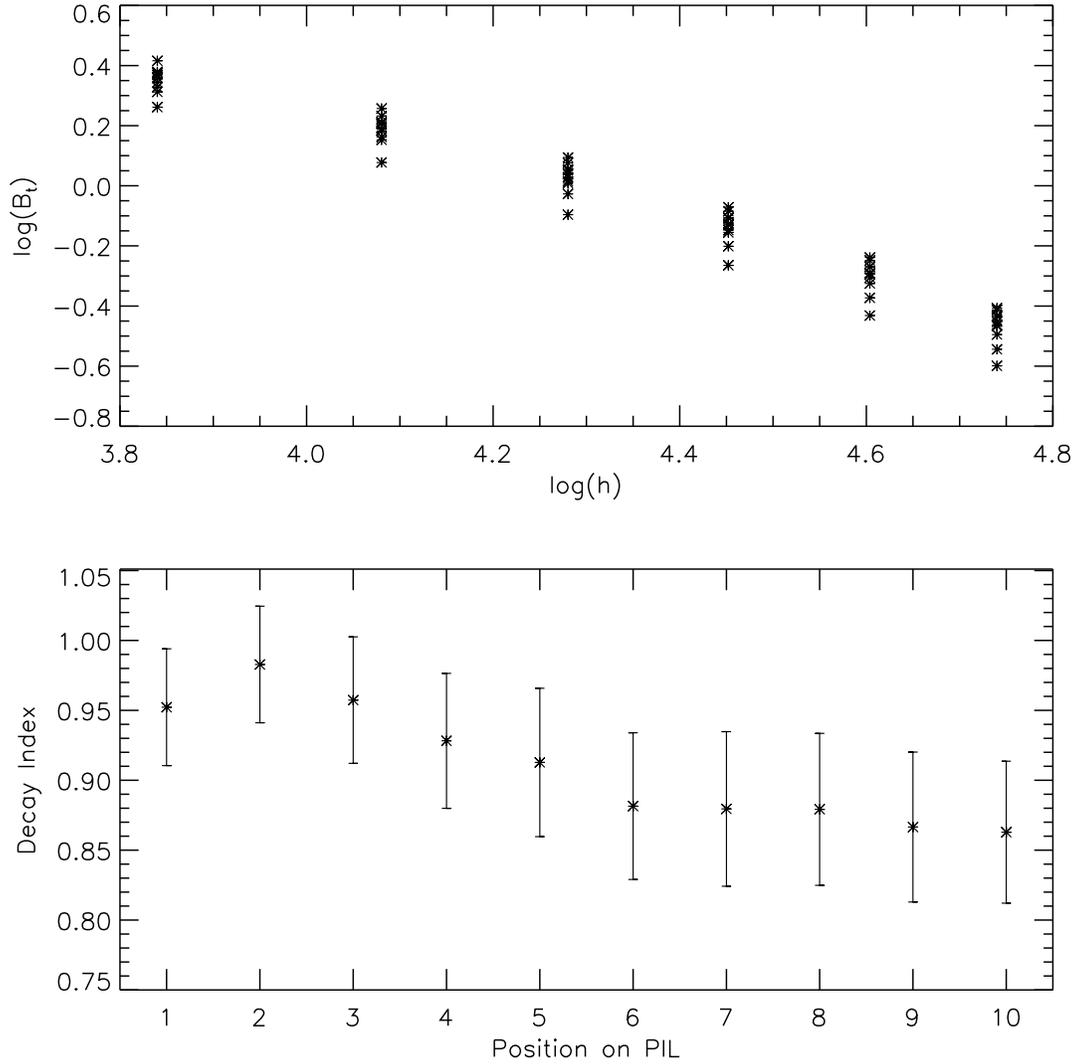}
\caption{ Upper panel: log(Bt) vs. log(h) along the PIL. A height range of 46.5 - 114.5 Mm was used from extrapolated fields to calculate the decay index. Lower panel: the decay indices derived from the extrapolated potential field along the PIL from the upper to the lower. On average, the decay index along the PIL is 0.91.}
\end{figure}

\begin{figure}
\epsscale{.90}
\plotone{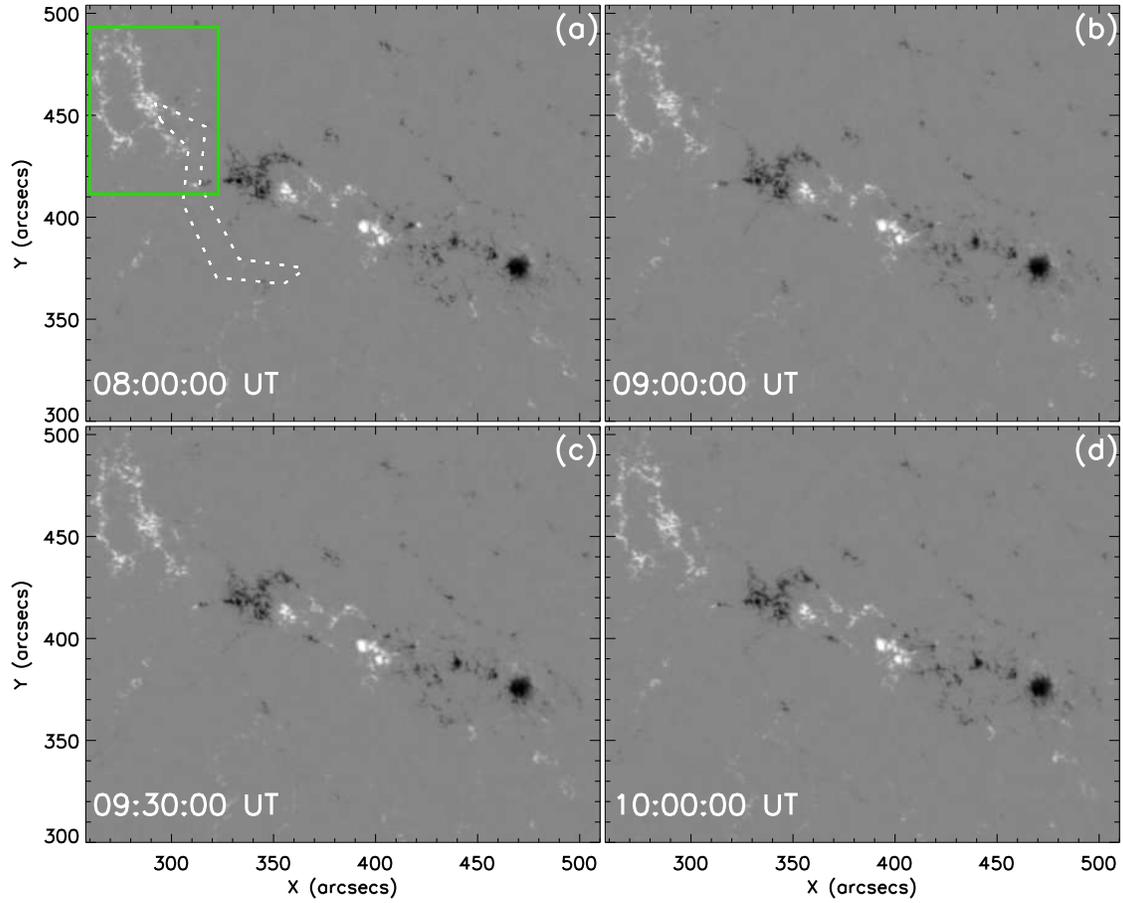}
\caption{Sequence of line-of-sight HMI magnetogram to show the evolution of the active region during the filament eruption. The green box is the same as the green box in Fig. 1, which is used to calculate the magnetic helicity at the foot-point of the filament. The dashed lines in Fig. 10a outline the boundary of the filament.\label{fig2}}
\end{figure}

\begin{figure}
\includegraphics[angle=0,scale=0.550]{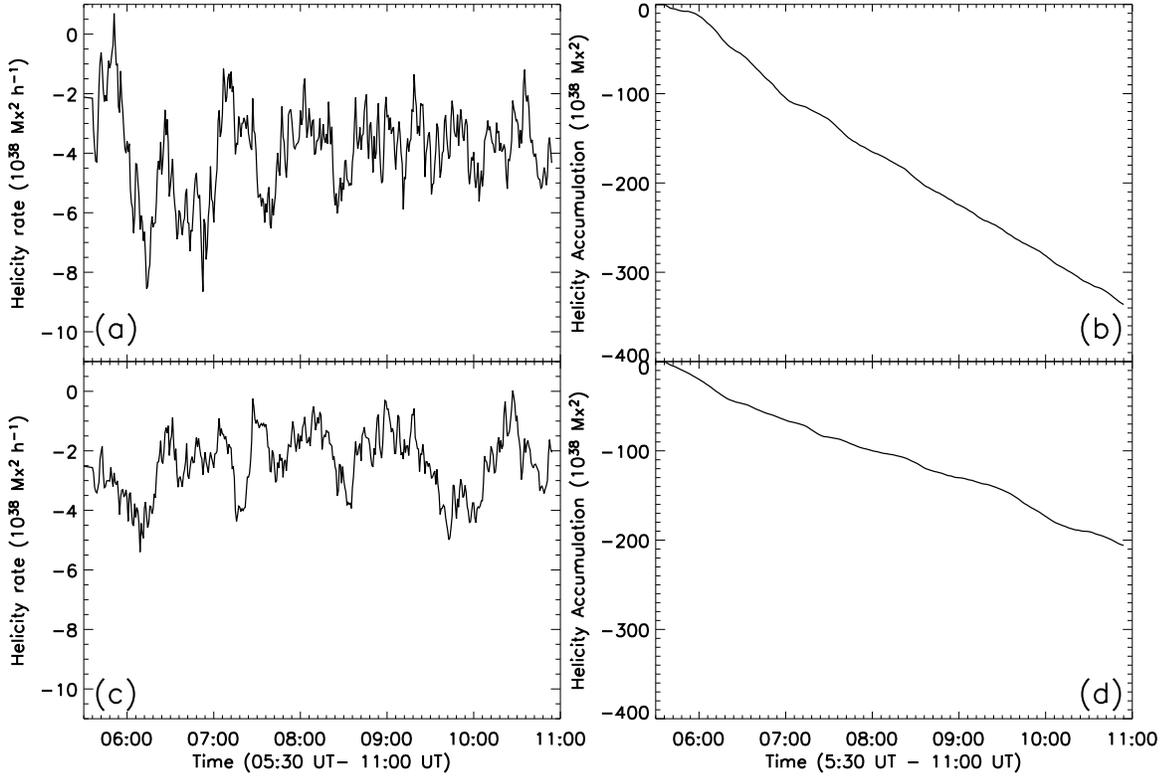}
\caption{The time profile of the helicity injection rate and the helicity accumulation by the horizontal motions. Figs. 11(a) and 11(b) show the time profile of the rate of the helicity injection and the helicity accumulation in the region marked by the green box in Fig. 10.  Figs. 11(c) and 11(d) show the time profile of the rate of the helicity injection and the helicity accumulation in the whole active region.}
\end{figure}

\end{document}